\documentclass[12pt]{iopart}
\usepackage{amssymb}
\usepackage{wrapfig}
\usepackage{graphicx}
\usepackage[usenames]{color}

\begin{document}
\title{Solutions to Maxwell's Equations using Spheroidal Coordinates}
\author{M. Zeppenfeld}
\address{Max-Planck-Institut f\"ur Quantenoptik, Hans-Kopfermann-Str.\,1, D-85748 Garching, Germany}
\ead{martin.zeppenfeld@mpq.mpg.de}

\begin{abstract}
Analytical solutions to the wave equation in spheroidal coordinates in the short wavelength limit are considered. The asymptotic solutions for the radial function are significantly simplified, allowing scalar spheroidal wave functions to be defined in a form which is directly reminiscent of the Laguerre-Gaussian solutions to the paraxial wave equation in optics. Expressions for the Cartesian derivatives of the scalar spheroidal wave functions are derived, leading to a new set of vector solutions to Maxwell's equations. The results are an ideal starting point for calculations of corrections to the paraxial approximation.
\end{abstract}


\section{Introduction}
Solutions to the wave equation in spheroidal coordinates have application to a wide range of problems in physics~\cite{Flammer57}. In particular, in the short wavelength limit, consideration of the wave equation in oblate spheroidal coordinates leads directly to the well known Gauss-Laguerre solutions to the paraxial approximation of the wave equation in optics~\cite{McDonald03}. The paraxial approximation is an extremely versatile tool for studying beams of coherent radiation (e.g. laser beams), and successfully describes phenomena such as propagation of beams through lens systems, focus size, wave front curvature and phase shifts in the focus of beams, as well as eigenmodes and eigenfrequencies of spherical mirror resonators~\cite{Siegman86}. Despite its great success, the paraxial approximation is only an approximation. Consideration of corrections beyond the paraxial approximation is relevant not only in order to establish a bound on its validity but is in fact necessary to describe strongly focused beams~\cite{Agrawal79} and to explain the fine structure of the frequency spectrum of high finesse resonators~\cite{Zeppenfeld09}.

Corrections to the paraxial approximation can be calculated by reconsidering the term which is neglected when deriving the paraxial approximation from the wave equation~\cite{Lax75}. As an alternative, corrections can be effectively considered using exact solutions to the wave equation in spheroidal coordinates. This has the advantage of removing ambiguities in the definition of higher-order terms, leads to simpler expressions, and is in general a more natural framework for exact consideration of beamlike solutions to the wave equation.

The wave equation is separable in spheroidal coordinates, allowing solutions to be written as the product of a so called radial function depending only on the $\xi$ coordinate, a so called angle function depending only on the $\eta$ coordinate and the function $e^{im\phi}$ depending only on the azimuthal $\phi$ coordinate~\cite{Flammer57}. For short wavelengths, solutions for the radial and angle functions in the form of asymptotic expansions have been known for a long time~\cite{Baber35,Svartholm38,Meixner44}. The close relationship between spheroidal coordinates and Laguerre-Gaussian beams suggests that the asymptotic solutions are in some way related to the Gauss-Laguerre solutions to the paraxial wave equation. While this relationship is directly obvious for the asymptotic solutions for the angle functions, the opposite is true for the previously known solutions for the radial functions. In fact, the previously known asymptotic solutions for the radial functions consist of multiple sums with a huge number of terms.

Using solutions to the wave equation for exact calculations in optics requires taking the transverse vector character of the electromagnetic fields into account. Full vector solutions to Maxwell's equations were already considered by Ref.~\cite{Flammer53}. These vector solutions are derived by applying vector operators to the scalar spheroidal solutions and are expressed as derivatives in the spheroidal coordinates of the scalar solutions. As for the asymptotic solutions for the radial functions, the resulting expressions are quite complicated.

In this paper, we derive a new expression for the asymptotic expansion of the radial functions which to lowest order is directly reminiscent of solutions of the paraxial approximation. The new expression is significantly simpler, making its analytic application practicable. In addition, we derive expressions for the Cartesian derivatives of the spheroidal wave functions which allow us to define a significantly simpler set of vector spheroidal wave functions. The component of the field transverse to the direction of propagation is equal to a single scalar spheroidal wave function. This creates a direct correspondence between our vector solutions and Gauss-Laguerre solutions to the paraxial approximation.

After a brief discussion on notation used in this paper in section~\ref{section notation}, we introduce oblate spheroidal coordinates in section~\ref{section coordinates}. The wave equation in spheroidal coordinates as well as separation of coordinates is discussed in section~\ref{section wave equation}. Section~\ref{section angle function} is a review of the asymptotic expansion of the angle functions. The derivation is equivalent to the one given e.g. in Ref.~\cite{Flammer57}. The radial functions are considered in section~\ref{section radial function}. Starting with the previously known asymptotic expansion allows us to derive a new expression for the two lowest-order terms. Taking the lowest-order term as an ansatz for the radial function leads to a differential equation which allows higher-order terms to be calculated iteratively. In section~\ref{section scalar wave function} we define scalar spheroidal wave functions and demonstrate their equivalence in the short wavelength limit to the Gauss-Laguerre solutions to the paraxial wave equation. This is followed by a brief discussion concerning the symmetry properties of the scalar spheroidal wave functions upon reflection through a plane containing the symmetry axis of the spheroidal coordinate system. We proceed in section~\ref{section scalar to vector} by deriving expressions for the Cartesian derivatives of the scalar spheroidal wave functions. These expressions allow us to define a new set of vector spheroidal wave functions in section~\ref{section vector wave function} which are transverse vector solutions to the wave equation. section~\ref{section vector wave function} concludes with expressions for the curl of the vector spheroidal wave functions.

\subsection{Notation}\label{section notation}
The notation used in this paper is generally consistent with the notation in Ref.~\cite{Flammer57}, with the exceptions noted here. Previously, the symbol $c$ has been used to quantify the scaling of the spheroidal coordinate system relative to the wavelength. Since it seems problematic to use the same symbol as for the speed of light in a theory with strong application to optics, we use $\bar{c}$ instead.

Additionally, Ref.~\cite{Flammer57} generally uses $m$ and $n$ to label the spheroidal functions and associated parameters, using the labels $m$ and $\nu$ with $(m,n)=(m,2\nu+m)$ only when Laguerre functions are directly involved. In the short wavelength limit, both sets of labels can be used. The advantage of the $m$, $\nu$ labeling is that it is identical to the labeling of the Laguerre polynomials, which are essential for the expansion of the angle functions in the short wavelength limit. As a result, we use the labels $m$ and $\nu$ exclusively.

Outside the short wavelength limit, not considered in this paper, use of the labels $m$ and $n$ is necessary. This is due to the fact that when the short wavelength limit does not apply, the labels $m$, $\nu$ do not uniquely label the set of solutions to the wave equation whereas the labels $m$, $n$ do. Each pair of values $m$, $\nu$ corresponds to two solutions for the angle function and two solutions for the radial function with opposite parity about the plane $\xi=0$. In the short wavelength limit, the two distinct angle functions converge to the same function asymptotically, and the two radial functions can be considered as the real and imaginary part of a single complex valued function, so that $m$, $\nu$ effectively labels all solutions uniquely.

In contrast to Ref.~\cite{Flammer57}, we attempt to consistently place the labels $m$ and $\nu$ as subscripts, generally placing other labels as superscripts. The only exception is the Laguerre functions, where we use the standard notation $L_\nu^{(m)}$. Since we are only interested in oblate spheroidal coordinates and real arguments $\eta$ and $\xi$, we write the angle and radial functions respectively as $S_{m\nu}(\eta)$ and $R_{m\nu}(\xi)$ instead of as $S_{mn}(-i\bar{c},\eta)$ and $R_{mn}(-i\bar{c},i\xi)$ as in Ref.~\cite{Flammer57}, thereby ignoring the relationship between solutions in oblate spheroidal coordinates and in prolate spheroidal coordinates. Finally, the indices for the sums over Laguerre functions as e.g. in Eq.~(\ref{laguerre expansion}) range from $0$ to $\infty$ rather than from $-\nu$ to $\infty$ as in Ref.~\cite{Flammer57}. This allows us to treat all solutions for the angle functions on an equal footing, without choosing a specific one in advance by fixing $\nu$ ahead of time.

\section{Spheroidal coordinates}\label{section coordinates}
\begin{wrapfigure}{r}{9cm}
\includegraphics[width=0.6\textwidth]{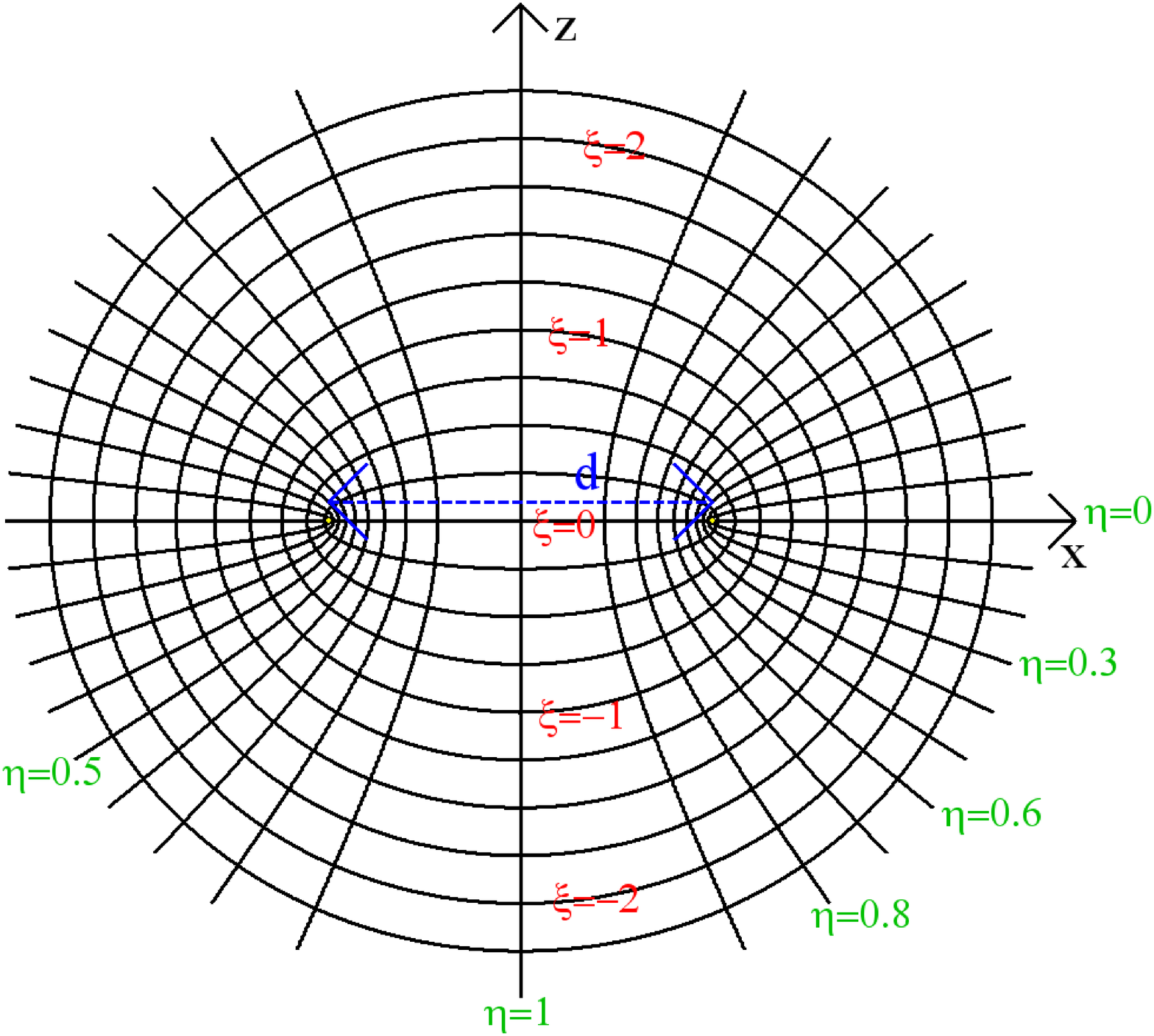}
\caption{Oblate Spheroidal Coordinates.}\label{Coordinates}
\end{wrapfigure}
The two-dimensional elliptic coordinate system is defined from the set of all ellipses and all hyperbolas with a common set of two focal points~\cite{Flammer57}. We denote the separation of the two focal points by $d$. Oblate spheroidal coordinates are derived from elliptic coordinates by rotating the elliptical coordinate system about the perpendicular bisector of the focal points~\cite{Flammer57}. The focal points thereby sweep out a circle of radius $d/2$. Oblate spheroidal coordinates are commonly mapped to Cartesian $(x,y,z)$ coordinates by placing this circle in the $x-y$ plane with center at the origin. The coordinates are often labeled $\eta$, $\xi$ and $\phi$ with the transformation to Cartesian coordinates given by~\cite{Flammer57}

\vspace{-5mm}
\begin{eqnarray}
\eqalign{x=\frac{d}{2}\sqrt{(1-\eta^2)(1+\xi^2)}\cos{\phi},\cr
y=\frac{d}{2}\sqrt{(1-\eta^2)(1+\xi^2)}\sin{\phi},\cr
\hspace{2cm}z=\frac{d}{2}\eta\xi,\cr
\hspace{-1cm}\eta\in[0,1],\hspace{.5cm}\xi\in(-\infty,\infty),\hspace{.5cm}\phi\in[0,2\pi).}
\end{eqnarray}
Oblate spheroidal coordinates reduced to the $x-z$ plane are shown in Fig. (\ref{Coordinates}). 

The coordinates $r$ and $\theta$ defined by
\begin{equation}\label{r and theta}
r=\sqrt{x^2+y^2}\hspace{1cm}\mbox{and}\hspace{1cm}\cos{\theta}=\eta
\end{equation}
are useful to describe aspects of the coordinate system. Note that near the $z$-axis, $\xi$ quantifies distance along the $z$-axis and $\theta$ quantifies distance from the  $z$-axis. To lowest order in $\theta$, $z=\frac{d}{2}\,\xi$ and $r=\frac{d}{2}\sqrt{1+\xi^2}\,\theta$.

\subsection{Wave equation in spheroidal coordinates}\label{section wave equation}
The wave equation,
\begin{equation}
(\nabla^2+k^2)\psi=0,
\end{equation}
written in oblate spheroidal coordinates, is given by~\cite{Flammer57}
\begin{eqnarray}\label{spheroidal wave eq}
\eqalign{\fl\left(\frac{\partial}{\partial\xi}(1+\xi^2)\frac{\partial}{\partial\xi}+\frac{\partial}{\partial\eta}(1-\eta^2)\frac{\partial}{\partial\eta}-\left(\frac{1}{1+\xi^2}-\frac{1}{1-\eta^2}\right)\frac{\partial^2}{\partial\phi^2}+\bar{c}^2(\eta^2+\xi^2)\right)\psi=0,\cr
\mbox{with}\hspace{.5cm}\bar{c}=\frac{kd}{2}.}
\end{eqnarray}
This equation is separable. Therefore $\psi$ can be written as a product of three functions depending only on $\eta$, $\xi$ and $\phi$, respectively. Due to cylindrical symmetry, the function depending on $\phi$ can be written as $e^{im\phi}$ for integer values of $m$,
\begin{equation}\label{function separation}
\psi=R_{m\nu}(\xi)S_{m\nu}(\eta)e^{im\phi}.
\end{equation}
The label $\nu$ is used to denote the various solutions of Eq.~(\ref{spheroidal wave eq}).

The so called radial function $R_{m\nu}(\xi)$ and angle function $S_{m\nu}(\eta)$ satisfy the two differential equations
\begin{equation}\label{radial eq}
\left(\frac{d}{d\xi}(1+\xi^2)\frac{d}{d\xi}+\frac{m^2}{1+\xi^2}+\bar{c}^2\xi^2-\lambda_{m\nu}\right)R_{m\nu}(\xi)=0
\end{equation}
and
\begin{equation}\label{angle eq}
\left(\frac{d}{d\eta}(1-\eta^2)\frac{d}{d\eta}-\frac{m^2}{1-\eta^2}+\bar{c}^2\eta^2+\lambda_{m\nu}\right)S_{m\nu}(\eta)=0,
\end{equation}
as can be seen from comparison with Eq.~(\ref{spheroidal wave eq}). Here, $\lambda_{m\nu}$ is the separation constant.

\section{Angle functions for large $\bar{c}$}\label{section angle function}
In the limit $\bar{c}\to\infty$, a solution to Eq.~(\ref{angle eq}) expressed as a series in $\frac{1}{\bar{c}}$ can be found as follows. The ansatz
\begin{equation}\label{angle ansatz}
S_{m}(\eta)=(1-\eta^2)^\frac{m}{2}e^{-\bar{c}(1-\eta)}s_{m}(x),\hspace{1cm}\mbox{with}\hspace{1cm}x=2\bar{c}(1-\eta),
\end{equation}
allows Eq.~(\ref{angle eq}) to be rewritten as
\begin{eqnarray}\label{radial rewrite}
\eqalign{\left[x\frac{d^2}{dx^2}+(m+1-x)\frac{d}{dx}-\frac{m+1}{2}+\frac{\bar{c}^2+\lambda_{m}}{4\bar{c}}+\right.\cr
\hspace{-15mm}\left.-\frac{x^2}{4\bar{c}}\frac{d^2}{dx^2}+\frac{x^2-2x(m+1)}{4\bar{c}}\frac{d}{dx}+\frac{x(m+1)-(m^2+m)}{4\bar{c}}\right]s_{m}(x)=0.}
\end{eqnarray}
We have dropped the index $\nu$ in order to refer to a general solution with arbitrary $\lambda_{m(\nu)}$. Note that by introducing the factor $(1-\eta^2)^\frac{m}{2}$ in Eq.~(\ref{angle ansatz}), we break the symmetry between positive and negative $m$. This symmetry nonetheless persists in the original Eq.~(\ref{angle eq}) so that the solutions must in the end be symmetric under a transformation from $m$ to $-m$. We address this issue in more detail in section~\ref{section scalar wave function}.

For $\bar{c}\to\infty$, the terms on the second line of Eq.~(\ref{radial rewrite}) inside the bracket vanish. What is left is simply the associated Laguerre differential equation. A possible set of approximate solutions to Eq.~(\ref{radial rewrite}) is therefore the set of associated Laguerre polynomials $L_\nu^{(m)}(x)$ with fixed $m$,
\begin{equation}\label{s_m,nu=L_m,nu}
s_{m}(x)=L_\nu^{(m)}(x)+{\cal O}(\frac{1}{\bar{c}}),
\end{equation}
and $\lambda_{m}$ given by
\begin{equation}\label{lambda_m}
\lambda_{m}=-\bar{c}^2+[2(m+1)+4\nu]\bar{c}+{\cal O}(1).
\end{equation}
The form of Eq.~(\ref{s_m,nu=L_m,nu}) suggests expanding the exact solution $s_{m}(x)$ in terms of associated Laguerre polynomials as
\begin{equation}\label{laguerre expansion}
s_{m}(x)=\sum_{s=0}^\infty A_{m}^sL_{s}^{(m)}(x).
\end{equation}
The $A_{m}^s$ are expansion coefficients. Inserting this expression into Eq.~(\ref{radial rewrite}) and using appropriate recursion relations for Laguerre polynomials allows the left hand side of Eq.~(\ref{radial rewrite}) to be rewritten as a linear combination of Laguerre polynomials with fixed $m$. Note that a number of useful recursion relations for Laguerre Polynomials are listed in \ref{appendix recursion}. Since the Laguerre polynomials with fixed $m$ are linearly independent, the coefficient of each term in this sum must be zero, leading to the following relations between the $A_{m}^s$,
\begin{eqnarray}\label{begin A_mnu beta_mnu}
\eqalign{\fl s(s+m)A_{m}^{s-1}+[(4s+2(m+1))\bar{c}-(2s^2+(2s+1)(m+1))-\lambda_{m}']A_{m}^s+\cr
(s+1)(s+m+1)A_{m}^{s+1}=0,\cr
\mbox{with}\hspace{1cm}\lambda_{m}'=\lambda_{m}+\bar{c}^2.}
\end{eqnarray}
These relations can be written in matrix form as
\begin{equation}\label{matrix recursion}
(\bar{c}M_0+M_1-\lambda_{m}'\mathbb{I})\mathbf{A}=\mathbf{0}.
\end{equation}
Here, $M_0$ is a diagonal matrix, $M_1$ is a tridiagonal matrix, $\mathbb{I}$ is the identity matrix and $\mathbf{A}$ is the vector of coefficients $A_{m}^s$ according to
\begin{eqnarray}
\eqalign{(M_0)_{s,t}=[4s+2(m+1)]\,\delta_{s,t},\cr
\fl(M_1)_{s,t}=s(s+m)\delta_{s-1,t}-(2s^2+(2s+1)(m+1))\delta_{s,t}+(s+1)(s+m+1)\delta_{s+1,t},\cr
(\mathbf{A})_s=A_{m}^s.}
\end{eqnarray}
We use the Kronecker delta function, $\delta_{s,t}$. Eq.~(\ref{matrix recursion}) transforms the problem of determining the coefficients in Eq.~(\ref{laguerre expansion}) into an eigenvalue problem. $\lambda_{m}'$ is an eigenvalue of $\bar{c}M_0+M_1$ and $\mathbf{A}$ is the corresponding eigenvector.

In the limit $\bar{c}\to\infty$, $M_1$ can be neglected compared to $\bar{c}M_0$. Since $M_0$ is diagonal, finding an approximate solution to Eq.~(\ref{matrix recursion}) is therefore trivial, this solution being given by Eqs.~(\ref{s_m,nu=L_m,nu}) and (\ref{lambda_m}). Improved approximate solutions can be found via perturbation theory~\cite{Sakurai94}. According to convention we reintroduce the label $\nu$ to denote the solution whose eigenvector is given to lowest order in $\frac{1}{\bar{c}}$ by $A_{m\nu}^s=\delta_{s,\nu}$. We expand $\lambda_{m\nu}'$ and $A_{m\nu}^s$ in powers of $\frac{1}{\bar{c}}$ as
\begin{eqnarray}\label{perturbation sum}
\eqalign{\lambda_{m\nu}'=\sum_{k=-1}^\infty\beta_{m\nu}^k\bar{c}^{-k},\cr
A_{m\nu}^s=\sum_{k=0}^\infty A_{m\nu}^{s,k}\bar{c}^{-k}.}
\end{eqnarray}
Inserting these expressions into Eq.~(\ref{matrix recursion}) and matching terms of equal order in $\frac{1}{\bar{c}}$ leads to recursive expressions for $\beta_{m\nu}^k$ and $A_{m\nu}^{s,k}$,
\begin{eqnarray}\label{angle coefficients}
\eqalign{\beta_{m\nu}^{-1}=(M_0)_{\nu,\nu}=4\nu+2(m+1),\cr
\beta_{m\nu}^k=\sum_s(M_1)_{\nu,s}A_{m\nu}^{s,k}&k\ge0,\cr
A_{m\nu}^{s,0}=\delta_{s,\nu},\hspace{1.5cm}A_{m\nu}^{\nu,k}=\delta_{k,0},\cr
A_{m\nu}^{s,k+1}=\frac{\sum_t(M_1)_{s,t}A_{m\nu}^{t,k}-\sum_{k'=0}^{k-1}\beta_{m\nu}^{k'}A_{m\nu}^{s,k-k'}}{(M_0)_{(\nu,\nu)}-(M_0)_{(s,s)}}\hspace{1cm}&k\ge0, s\ne\nu.}
\end{eqnarray}
Since $M_1$ is tridiagonal, $A_{m\nu}^{s,k}$ is zero for $|s-\nu|>k$. As a result, each sum over $s$ or $t$ contains only a finite number of nonzero terms.

Eqs.~(\ref{angle coefficients}) lead to the expressions for $\beta_{m\nu}^k$ and $A_{m\nu}^{s,k}$ for $k\le4$ given explicitly in Ref.~\cite{Flammer57}. Higher-order values can easily be obtained numerically from Eqs.~(\ref{angle coefficients}). Note that the expression for $\beta_4^{mn}$ in Ref.~\cite{Flammer57} contains an incorrect "$-$" sign.

\section{Radial functions for large $\bar{c}$}\label{section radial function}
Following a similar procedure as the one in section~\ref{section angle function}, it is also possible to find an asymptotic expansion for the radial function, given by \cite{Flammer57}
\begin{equation}\label{old radial}
\hspace{-1cm}R_{m\nu}(\xi)=\frac{2\pi(2\bar{c})^mi^{-(2\nu+1)}}{m!A_{m\nu}^0}e^{-\bar{c}(2+i\xi)}(\xi^2+1)^{m/2}\sum_{s=0}^\infty A_{m\nu}^sU_s^{(m)}[2\bar{c}(1+i\xi)].
\end{equation}
Note that Ref.~\cite{Flammer57} defines a total of four different radial functions $R_{mn}^{(j)}(-i\,\bar{c},i\,\xi)$ with $j=1,...,4$. These are respectively the real part of Eq.~(\ref{old radial}), the imaginary part of Eq.~(\ref{old radial}), Eq.~(\ref{old radial}) itself and the complex conjugate of Eq.~(\ref{old radial}). Without loss of generality, we only consider Eq.~(\ref{old radial}). The $A_{m\nu}^s$ in Eq.~(\ref{old radial}) are the expansion coefficients for the angle function from Eq.~(\ref{laguerre expansion}) which can be calculated using Eqs.~(\ref{perturbation sum}) and (\ref{angle coefficients}). The $U_\nu^{(m)}(z)$ are second solutions of the Laguerre differential equation, defined by~\cite{Pinney46}
\begin{equation}\label{second laguerre}
U_\nu^{(m)}(z)=i\,\frac{e^{-im\pi}L_\nu^{(m)}(z)-\frac{\Gamma(\nu+m+1)}{\Gamma(\nu+1)}z^{-m}L_{\nu+m}^{(-m)}(z)}{\sin(m\pi)}.
\end{equation}
For integer values of $m$, the denominator and numerator in Eq.~(\ref{second laguerre}) both equal zero, in which case $U_\nu^{(m)}(z)$ is defined as a limit in $m$. For large $|z|$, an asymptotic expansion for $U_\nu^{(m)}(z)$ exists, given by~\cite{Flammer57,Pinney46}
\begin{equation}\label{second laguerre expansion}
U_\nu^{(m)}(z)=\frac{e^{i(\nu+1/2)\pi}e^z}{\nu!\,\pi\, z^{\nu+m+1}}\sum_{r=0}^\infty\frac{(\nu+r)!\,(\nu+m+r)!}{r!\,z^r}.
\end{equation}

Unlike the asymptotic expansion for the angle function, which is quite transparent since it contains a single Laguerre polynomial to lowest order and additional Laguerre polynomials as higher-order corrections, the asymptotic expansion for the radial function is rather opaque since the terms $s=0$ through $s=\nu$ all contain terms of equal and lowest order in $\frac{1}{\bar{c}}$ as can be seen from Eqs.~(\ref{old radial}) and (\ref{second laguerre expansion}). We attempt to rewrite Eq.~(\ref{old radial}) as a sum over terms of equal order in $\frac{1}{\bar{c}}$. To this end we insert the asymptotic expansions for $U_\nu^{(m)}(z)$ and $A_{m\nu}^s$, Eqs.~(\ref{second laguerre expansion}) and (\ref{perturbation sum}), into Eq.~(\ref{old radial}). This leads to sums over $r$, $s$ and $k$. We define $p=r+s+k-\nu$ and observe that the total order in $\frac{1}{\bar{c}}$ of a term in the triple sum is $p+1$. Replacing the sum over $k$ by a sum over $p$, one obtains
\begin{equation}\label{many sums}
\fl R_{m\nu}(\xi)=\sum_{p=0}^\infty\sum_{r=0}^p\sum_{s=0}^{\nu+\lfloor\frac{p-r}{2}\rfloor}\frac{(-1)^{s-\nu}}{2^{r+s}}\frac{(r+s)!(m+r+s)!}{m!\,r!\,s!}\bar{c}^{-(p+\nu+1)}\frac{A_{m\nu}^{s,k}}{A_{m\nu}^0}e^{i\bar{c}\xi}\frac{(\xi^2+1)^{m/2}}{(1+i\xi)^{m+r+s+1}}.\hspace{4mm}
\end{equation}
The symbol $\lfloor.\rfloor$ denotes rounding down to the nearest integer.

As we will demonstrate, the sums over $r$ and $s$ can be performed resulting in relatively compact analytic expressions for all values of $p$. We start by considering the two lowest-order cases, $p=0$ and $p=1$. To proceed, we need analytic expressions for $A_{m\nu}^{s,k}$ for $s+k\le\nu+1$. Using Eqs.~(\ref{angle coefficients}), it can be shown by induction that
\begin{equation}
A_{m\nu}^{s,\nu-s}=2^{-2(\nu-s)}\frac{\nu!(\nu+m)!}{(\nu-s)!s!(m+s)!},
\end{equation}
and
\begin{equation}
A_{m\nu}^{s,\nu-s+1}=2^{-2(\nu-s+1)}\frac{(3\nu+2m+s+1)\nu!(\nu+m)!}{(\nu-s-1)!s!(m+s)!}.
\end{equation}
For $p=0$, one obtains
\begin{eqnarray}\label{p=0}
\eqalign{\sum_{s=0}^\nu\frac{(-1)^{s-\nu}}{2^s}\frac{(m+s)!}{m!}\bar{c}^{-(\nu+1)}\frac{A_{m\nu}^{s,\nu-s}}{A_{m\nu}^0}e^{i\bar{c}\xi}\frac{(\xi^2+1)^{m/2}}{(1+i\xi)^{m+s+1}}=\cr
\hspace{1cm}\frac{\bar{c}^{-(\nu+1)}}{A_{m\nu}^0}\frac{(\nu+m)!}{2^{2\nu}m!}e^{i\bar{c}\xi}\frac{(1-i\xi)^{\nu+m/2}}{(1+i\xi)^{\nu+m/2+1}}.}
\end{eqnarray}
The first line is Eq.~(\ref{many sums}) with fixed $p=0$. To obtain the second line, define $z=1+i\xi$ and observe that the first line is essentially the binomial expansion of $(1-\frac{z}{2})^\nu$.

The case $p=1$ is already slightly more complicated,
\begin{eqnarray}\label{p=1}
\eqalign{\sum_{r=0}^1\sum_{s=0}^\nu\frac{(-1)^{s-\nu}}{2^{r+s}}\frac{(r+s)!(m+r+s)!}{m!\,s!}\bar{c}^{-(\nu+2)}\frac{A_{m\nu}^{s,\nu+1-r-s}}{A_{m\nu}^0}e^{i\bar{c}\xi}\frac{(\xi^2+1)^{m/2}}{(1+i\xi)^{m+r+s+1}}=\hspace{1cm}\cr
\fl\frac{\bar{c}^{-(\nu+2)}}{A_{m\nu}^0}\frac{(\nu+m)!}{2^{2\nu}m!}\left[\frac{(\nu+1)(\nu+m+1)}{2(1+i\xi)}+\frac{3\nu^2+2m\nu+\nu}{4}-\frac{\nu(\nu+m)}{2(1-i\xi)}\right]e^{i\bar{c}\xi}\frac{(1-i\xi)^{\nu+m/2}}{(1+i\xi)^{\nu+m/2+1}}.}
\end{eqnarray}
The first line is again Eq.~(\ref{many sums}) but with fixed $p=1$. The second line is obtained in a similar manner as before, by observing that for $r=0$ the first line is essentially the binomial expansion of
$z^{3\nu+m+1}\frac{d}{dz}\,z^{-(4\nu+2m)}\left(1-\frac{z}{2}\right)^{\nu-1},$
and for $r=1$ the first line is essentially the binomial expansion of
$\frac{d}{dz}\,z^{-(m-1)}\frac{d}{dz}\,z^{-(\nu+1)}\left(1-\frac{z}{2}\right)^\nu.$

Using Eqs.~(\ref{p=0}) and (\ref{p=1}), we obtain a preliminary result for the summation of terms of equal order in $\frac{1}{\bar{c}}$ for the asymptotic expansion of $R_{m\nu}(\xi)$ through first order in $\frac{1}{\bar{c}}$. In writing the preliminary result, we must confront the issue that the higher-order terms are to some degree arbitrary. $R_{m\nu}(\xi)$ is defined by its differential equation only up to some constant factor. In particular, we can multiply $R_{m\nu}(\xi)$ by a constant which depends on $\bar{c}$. The asymptotic expansion in $\frac{1}{\bar{c}}$ of such a product contains different higher-order terms. Specifically, a constant $a$ times the zeroth order term can be added to the $p$\,th order term as long as $a$ times the $n$th order term is simultaneously added to the $(p+n)$th order term, resulting in a different asymptotic expansion. This issue is raised by the presence of $A_{m\nu}^0$ as part of the normalization constant in Eq.~(\ref{old radial}), which depends on $\frac{1}{\bar{c}}$ through first order according to
\begin{equation}\label{A_mnu0}
\hspace{-1cm}A_{m\nu}^0=\bar{c}^{-\nu}2^{-2\nu}\frac{(\nu+m)!}{m!}+\bar{c}^{-\nu-1}2^{-2\nu}\frac{(3\nu+2m+1)\nu(\nu+m)!}{4m!}+{\cal O}(\bar{c}^{-\nu-2}).\hspace{4mm}
\end{equation}
Replacing $A_{m\nu}^0$ according to Eq.~(\ref{A_mnu0}) leads to
\begin{equation}\label{initial new radial}
\fl R_{m\nu}(\xi)=\bar{c}^{-1}e^{i\bar{c}\xi}\frac{(1-i\xi)^{\nu+m/2}}{(1+i\xi)^{\nu+m/2+1}}\left[1+\frac{1}{\bar{c}}\left(\frac{(\nu+1)(\nu+m+1)}{2(1+i\xi)}-\frac{\nu(\nu+m)}{2(1-i\xi)}\right)+{\cal O}\left(\frac{1}{\bar{c}^2}\right)\right].\hspace{4mm}
\end{equation}
This is a new and significantly simpler expression for the asymptotic expansion of the radial function through first order in $\frac{1}{\bar{c}}$.

Continuing the previous procedure for larger values of $p$ becomes increasingly cumbersome, not least because analytic expressions for $A_{m\nu}^{s,k}$ for larger values of $s+k-\nu$ become increasingly difficult to find. The similar form of the results for $p=0$ and $p=1$ suggests an alternative approach. We factor out the $\xi$ dependent terms in front of the bracket in Eq.~(\ref{initial new radial}),
\begin{equation}\label{new radial}
R_{m\nu}(\xi)=e^{i\bar{c}\xi}\frac{(1-i\xi)^{\nu+m/2}}{(1+i\xi)^{\nu+m/2+1}}\,r_{m\nu}(\xi).
\end{equation}
Inserting this expression into the differential equation for $R_{m\nu}(\xi)$ leads to a differential equation for $r_{m\nu}(\xi)$,
\begin{eqnarray}\label{r diffeq}
\eqalign{\left[(1+\xi^2)\frac{d^2}{d\xi^2}+i\,(2\bar{c}(1+\xi^2)-2(2\nu+m+1))\frac{d}{d\xi}+\right.\cr
\left.-\frac{2(\nu+1)(\nu+m+1)}{1+i\xi}-\frac{2\nu(\nu+m)}{1-i\xi}-\lambda_{m\nu}''\right]r_{m\nu}(\xi)=0,\cr
\mbox{with}\hspace{1cm}\lambda_{m\nu}''=\lambda_{m\nu}+\bar{c}^2-[2(m+1)+4\nu]\bar{c}.}
\end{eqnarray}
We can write $r_{m\nu}(\xi)$ as an asymptotic expansion in $\frac{1}{\bar{c}}$,
\begin{equation}
r_{m\nu}(\xi)=\sum_{p=0}^\infty \bar{c}^{-p}\,r_{m\nu}^{(p)}(\xi).
\end{equation}
Eq.~(\ref{r diffeq}) can be transformed into a recursive expression for the $r_{m\nu}^{(p)}(\xi)$,
\begin{eqnarray}\label{r recursion}
\eqalign{r_{m\nu}^{(p+1)}(\xi)=\int\frac{i}{2(1+\xi^2)}\left[\left((1+\xi^2)\frac{d^2}{d\xi^2}-2i(2\nu+m+1)\frac{d}{d\xi}+\right.\right.\cr
\hspace{-5mm}\left.\left.-\frac{2(\nu+1)(\nu+m+1)}{1+i\xi}-\frac{2\nu(\nu+m)}{1-i\xi}\right)r_{m\nu}^{(p)}(\xi)-\sum_{k=0}^p\beta_{m\nu}^kr_{m\nu}^{(p-k)}(\xi)\right]d\xi.}
\end{eqnarray}
The $\beta_{m\nu}^k$ are the expansion coefficients of $\lambda_{m\nu}$ according to Eq.~(\ref{perturbation sum}). Through $p=2$ one obtains
\begin{eqnarray}\label{r_mnu}
\eqalign{r_{m\nu}^{(0)}(\xi)=1,\hspace{2cm}
r_{m\nu}^{(1)}(\xi)=\frac{(\nu+1)(\nu+m+1)}{2(1+i\xi)}-\frac{\nu(\nu+m)}{2(1-i\xi)},\cr
\fl r_{m\nu}^{(2)}(\xi)=\frac{1}{8}\left(\frac{(\nu+1)(\nu+2)(\nu+m+1)(\nu+m+2)}{(1+i\xi)^2}+\frac{\nu(\nu-1)(\nu+m)(\nu+m-1)}{(1-i\xi)^2}-\right.\hspace{1cm}\cr
\fl\left.\frac{(\nu+1)(\nu+m+1)(\nu^2+m\nu-2\nu-m-1)}{1+i\xi}-\frac{\nu(\nu+m)(\nu^2+m\nu+4\nu+2m+2)}{1-i\xi}\right).}
\end{eqnarray}
The $r_{m\nu}^{(p)}$ through $p=4$ are listed in \ref{appendix r_mnu^p}.

Note that the $r_{m\nu}^{(p)}(\xi)$ are only defined up to a constant of integration. This reflects the discussion before Eq.~(\ref{A_mnu0}). Changing the constant of integration for $r_{m\nu}^{(p)}(\xi)$ is equivalent to adding $r_{m\nu}^{(0)}(\xi)$ and changing the higher-order functions accordingly as discussed above. We stick to the convention resulting in the presence of $A_{m\nu}^0$ in the normalization coefficient of Eq.~(\ref{old radial}) that $\lim_{|\xi|\to\infty}r_{m\nu}^{(p)}(\xi)=0$ for $p>0$.

Eq.~(\ref{r recursion}) provides a systematic method for determining analytic expressions for the summation over $r$ and $s$ in Eq.~(\ref{many sums}) as claimed directly thereafter. The $R_{m\nu}(\xi)$ defined by Eq.~(\ref{old radial}) and the $R_{m\nu}(\xi)$ defined by Eq.~(\ref{new radial}) are solutions to the same differential equation with the same boundary conditions and are therefore equal, up to a constant factor. The asymptotic expansions of both functions are therefore equivalent so that the summations over $r$ and $s$ must be obtainable from the $r_{m\nu}^{(p)}(\xi)$.

\section{Scalar spheroidal wave functions}\label{section scalar wave function}
Following Eq.~(\ref{function separation}), we define the scalar spheroidal wave functions $\psi_{m\nu}$ according to
\begin{equation}\label{spheroidal wave function}
\fl\psi_{m\nu}(\phi,\eta,\xi)=e^{im\phi}\times\left(\bar{c}^{m/2}(1-\eta^2)^{m/2}e^{-\bar{c}(1-\eta)}s_{m\nu}(\eta)\right)\times\left(\frac{(1-i\xi)^{\nu+m/2}}{(1+i\xi)^{\nu+m/2+1}}\,e^{i\bar{c}\xi}\,r_{m\nu}(\xi)\right).\hspace{4mm}
\end{equation}
Note the presence of the factor $\bar{c}^{m/2}$ which is included because $1-\eta^2\sim{\cal O}(\frac{1}{\bar{c}})$ (see below) and therefore $\bar{c}^{m/2}(1-\eta^2)^{m/2}\sim{\cal O}(1)$.

We briefly demonstrate the equality to lowest order in $\frac{1}{\bar{c}}$ of Eq.~(\ref{spheroidal wave function}) to the Gauss-Laguerre solutions to the paraxial wave equation. According to Eq.~(\ref{r and theta}) we have $1-\eta\sim\theta^2/2$. Due to the presence of the term $e^{-\bar{c}(1-\eta)}$, $\psi_{m\nu}$ takes on finite values only for $\theta={\cal O}(\bar{c}^{-1/2})$. We can therefore make the following substitutions in Eq.~(\ref{spheroidal wave function}).
$$(1-\eta^2)=\theta^2+{\cal O}\left(\frac{1}{\bar{c}^2}\right),\hspace{2cm}
\xi=\frac{2z}{d}\left(1+\frac{\theta^2}{2}+{\cal O}\left(\frac{1}{\bar{c}^2}\right)\right),$$
$$s_{m\nu}(\eta)=L_\nu^{(m)}(2\bar{c}(1-\eta))+{\cal O}\left(\frac{1}{\bar{c}}\right)=L_\nu^{(m)}(\bar{c}\,\theta^2)+{\cal O}\left(\frac{1}{\bar{c}}\right),$$
$$r_{m\nu}(\xi)=1+{\cal O}\left(\frac{1}{\bar{c}}\right),\hspace{2cm}
\theta^2=\frac{4r^2}{d^2(1+\xi^2)}+{\cal O}\left(\frac{1}{\bar{c}^2}\right)=\frac{4r^2}{d^2+4z^2}+{\cal O}\left(\frac{1}{\bar{c}^2}\right),$$
so that
\begin{eqnarray}
\eqalign{\fl\psi_{m\nu}=e^{im\phi}\times\left((\bar{c}\,\theta^2)^{m/2}e^{-\bar{c}\,\theta^2/2}L_\nu^{(m)}(\bar{c}\,\theta^2)\right)\times\left(\frac{(1-2iz/d)^{\nu+m/2}}{(1+2iz/d)^{\nu+m/2+1}}\,e^{2i\bar{c}z(1+\theta^2/2)/d}\right)+{\cal O}\left(\frac{1}{\bar{c}}\right)\cr
\fl\hspace{8mm} =e^{im\phi}\times\left(\left(\frac{2\,k\,d\,r^2}{d^2+4z^2}\right)^{m/2}L_\nu^{(m)}\left(\frac{2\,k\,d\,r^2}{d^2+4z^2}\right)e^{\frac{ikr^2}{2z-id}}\frac{(1-2iz/d)^{\nu+m/2}}{(1+2iz/d)^{\nu+m/2+1}}\,e^{ikz}\right)+{\cal O}\left(\frac{1}{\bar{c}}\right).\hspace{1cm}}
\end{eqnarray}
This is the equation for a Gauss-Laguerre beam with confocal parameter $d$~\cite{Siegman86}.

While the differential equations for the spheroidal functions are independent of the sign of $m$, the approach used to find the asymptotic series for the spheroidal functions introduces a dependence on the sign of $m$. As a result, it is by no means obvious that there is some symmetry between positive and negative $m$ in Eq.~(\ref{spheroidal wave function}). Demonstrating this symmetry is in fact nontrivial. The fact that the Laguerre polynomials transform from positive to negative $m$ according to~\cite{Abramowitz+Stegun}
\begin{equation}\label{m to -m}
L_{\nu+m}^{(-m)}(x)=\frac{\nu!}{(\nu+m)!}(-x)^mL_\nu^{(m)}(x)
\end{equation}
suggests that the spheroidal wave functions remain invariant under the transformation $(m,\nu)\to(-m,\nu+m)$, except possibly for a constant factor. While Eqs.~(\ref{new radial}-\ref{r_mnu}) involved in the new asymptotic expansion for the radial functions as well as Eqs.~(\ref{begin A_mnu beta_mnu}-\ref{angle coefficients}) involving the $A_{m(\nu)}^{s(,k)}$ and $\beta_{m\nu}^k$ are all seen to remain invariant under this transformation (i.e. $(m,\nu,s)\to(-m,\nu+m,s+m)$), complications arise for the angle functions due to the factor $(1-\eta^2)^{m/2}$ in the asymptotic expansion of $S_{m\nu}$ and due to the dependence of Eq.~(\ref{m to -m}) on $\nu$ and $m$. These two effects in fact cancel as we checked through first order in $\frac{1}{\bar{c}}$, leading to
\begin{equation}
\fl\psi_{-m,\nu+m}(-\phi,\eta,\xi)=(-1)^m\frac{\nu!}{(\nu+m)!}\left(1+\frac{m(2\nu+m+1)}{4\bar{c}}+{\cal O}\left(\frac{1}{\bar{c}^2}\right)\right)\psi_{m\nu}(\phi,\eta,\xi).\hspace{4mm}
\end{equation}
Surprisingly the transformation from positive to negative $m$ involves a factor which depends on $\bar{c}$.

\section{From scalar to vector wave functions}\label{section scalar to vector}
The scalar spheroidal wave functions defined by Eq.~(\ref{spheroidal wave function}) are solutions to the wave equation. Nonetheless, they are not suitable for exact calculations in optics, since solutions to Maxwell's equations are vector functions. Let $\mathbf{E}$ be the electric field of a solution to Maxwell's equations. In order to specify $\mathbf{E}$, it is useful to choose a vector basis and to specify the components of $\mathbf{E}$ in this basis. Since our wave functions are defined in spheroidal coordinates, it seems reasonable to express $\mathbf{E}$ in terms of the spheroidal unit vectors, $\mathbf{E}=E_\xi\mathbf{\hat{e}_\xi}+E_\eta\mathbf{\hat{e}_\eta}+E_\phi\mathbf{\hat{e}_\phi}$. Doing so has the major disadvantage that the individual components $E_\xi$, $E_\eta$ and $E_\phi$ are not solutions to the wave equation and therefore have no obvious relationship to the scalar spheroidal wave functions. We avoid this problem by using Cartesian unit vectors and write $\mathbf{E}=E_x\mathbf{\hat{x}}+E_y\mathbf{\hat{y}}+E_z\mathbf{\hat{z}}$.

A sufficient condition for $\mathbf{E}$ to be the electric field of a solution to Maxwell's equations in free space is that each of the components $E_x$, $E_y$ and $E_z$ satisfies the scalar wave equation and that $\nabla\cdot\mathbf{E}=0$~\cite{Jackson98}. As a result, we can construct solutions for $\mathbf{E}$ from our scalar spheroidal wave functions by writing each of $E_x$, $E_y$ and $E_z$ as a linear combination of the $\psi_{m\nu}$ such that $\nabla\cdot\mathbf{E}=0$ is satisfied. Since $\nabla\cdot\mathbf{E}=\frac{\partial E_x}{\partial x}+\frac{\partial E_y}{\partial y}+\frac{\partial E_z}{\partial z}$, satisifying $\nabla\cdot\mathbf{E}=0$ requires knowledge of $\frac{\partial\psi_{m\nu}}{\partial x}$, $\frac{\partial\psi_{m\nu}}{\partial y}$ and $\frac{\partial\psi_{m\nu}}{\partial z}$. Since Cartesian derivatives commute with the wave equation, $\frac{\partial\psi_{m\nu}}{\partial x}$, $\frac{\partial\psi_{m\nu}}{\partial y}$ and $\frac{\partial\psi_{m\nu}}{\partial z}$ are again solutions to the wave equation, and can therefore be written as linear combinations of the $\psi_{m\nu}$. We try to determine the coefficients of these linear combinations.

We begin by deriving expressions for $\frac{\partial}{\partial x}$, $\frac{\partial}{\partial y}$ and $\frac{\partial}{\partial z}$ in spheroidal coordinates. The scale factors in spheroidal coordinates, defined by  $h_{u_i}=\left|\frac{\partial\mathbf{x}}{\partial u_i}\right|$ with $u_i\in\{\xi,\eta,\phi\}$, are~\cite{Flammer57}
\begin{equation}
\fl h_\xi=\frac{d}{2}\sqrt{\frac{\eta^2+\xi^2}{1+\xi^2}},\hspace{1cm}
h_\eta=\frac{d}{2}\sqrt{\frac{\eta^2+\xi^2}{1-\eta^2}},\hspace{1cm}
h_\phi=\frac{d}{2}\sqrt{(1-\eta^2)(1+\xi^2)}.
\end{equation}
The spheroidal unit vectors, given by $\mathbf{\hat{e}}_{u_i}=\frac{1}{h_{u_i}}\frac{\partial\mathbf{x}}{\partial u_i}$ and expressed in terms of their Cartesian components, are
\begin{eqnarray}\label{unit vectors}
\eqalign{\mathbf{\hat{e}}_\xi=\left(\xi\sqrt{\frac{1-\eta^2}{\eta^2+\xi^2}}\right.\left.\cos\phi,\xi\sqrt{\frac{1-\eta^2}{\eta^2+\xi^2}}\sin\phi,\eta\sqrt{\frac{1+\xi^2}{\eta^2+\xi^2}}\right),\cr
\mathbf{\hat{e}}_\eta=\left(-\eta\sqrt{\frac{1+\xi^2}{\eta^2+\xi^2}}\right.\left.\cos\phi,-\eta\sqrt{\frac{1+\xi^2}{\eta^2+\xi^2}}\sin\phi,\xi\sqrt{\frac{1-\eta^2}{\eta^2+\xi^2}}\right),\cr
\mathbf{\hat{e}}_\phi=\left(-\sin\phi,\cos\phi,0\right).}
\end{eqnarray}
The gradient of a function expressed in spheroidal coordinates is
\begin{equation}\label{gradient}
\fl\nabla\psi=\sum_i\frac{\mathbf{\hat{e}}_{u_i}}{h_{u_i}}\frac{\partial\psi}{\partial u_i}=\frac{2}{d}\left(\mathbf{\hat{e}}_\xi\sqrt{\frac{1+\xi^2}{\eta^2+\xi^2}}\frac{\partial\psi}{\partial\xi}+\mathbf{\hat{e}}_\eta\sqrt{\frac{1-\eta^2}{\eta^2+\xi^2}}\frac{\partial\psi}{\partial\eta}+\mathbf{\hat{e}}_\phi\frac{1}{\sqrt{(1-\eta^2)(1+\xi^2)}}\frac{\partial\psi}{\partial\phi}\right).\hspace{4mm}
\end{equation}
Obtaining expressions for $\frac{\partial}{\partial x}$, $\frac{\partial}{\partial y}$ and $\frac{\partial}{\partial z}$ is simply a matter of combining Eqs.~(\ref{unit vectors}) and (\ref{gradient}),
\begin{equation}\label{d_dx+d_dy}
\fl\frac{\partial}{\partial x}\pm i\frac{\partial}{\partial y}=\frac{2}{d}\sqrt{(1-\eta^2)(1+\xi^2)}e^{\pm i\phi}\left(\frac{\xi}{\eta^2+\xi^2}\frac{\partial}{\partial\xi}-\frac{\eta}{\eta^2+\xi^2}\frac{\partial}{\partial\eta}\pm\frac{i}{(1-\eta^2)(1+\xi^2)}\frac{\partial}{\partial\phi}\right),\hspace{4mm}
\end{equation}
and
\begin{equation}\label{d_dz}
\frac{\partial}{\partial z}=\frac{2}{d}\left(\frac{\eta(1+\xi^2)}{\eta^2+\xi^2}\frac{\partial}{\partial\xi}+\frac{\xi(1-\eta^2)}{\eta^2+\xi^2}\frac{\partial}{\partial\eta}\right).
\end{equation}
From here on, one can obtain expressions for the Cartesian derivatives of the $\psi_{m\nu}$ essentially by brute force. Eqs.~(\ref{d_dx+d_dy}) and (\ref{d_dz}) are applied to Eq.~(\ref{spheroidal wave function}). The result is compared to successive orders in $\frac{1}{\bar{c}}$ to Eq.~(\ref{spheroidal wave function}). One obtains,
\begin{eqnarray}\label{dpsi_dx+idpsi_dy}
\eqalign{\left(\frac{\partial}{\partial x}+i\frac{\partial}{\partial y}\right)\psi_{m\nu}=\frac{-2\sqrt{\bar{c}}}{d}\left[\psi_{m+1,\nu-1}+\psi_{m+1,\nu}-{\color{White}\frac{0}{1}}\right.\cr
\hspace{-15mm}\frac{1}{\bar{c}}\left(\frac{\nu+m}{4}\psi_{m+1,\nu-2}-\frac{\nu}{4}\,\psi_{m+1,\nu-1}-\frac{\nu+m+1}{4}\psi_{m+1,\nu}+\frac{\nu+1}{4}\psi_{m+1,\nu+1}\right)+\cr
\hspace{-1cm}\frac{1}{\bar{c}^2}\left(\frac{(\nu+m)(\nu+m-1)}{16}\psi_{m+1,\nu-3}-\frac{(\nu+m)(3\nu+m-1)}{8}\psi_{m+1,\nu-2}+\right.\cr
\frac{(2\nu^3+4m\nu ^2+7\nu^2+2m^2\nu+4m\nu +3\nu-m^2)}{16}\psi_{m+1,\nu-1}-\cr
\frac{(2\nu^3+2m\nu ^2-\nu^2-6m\nu-5\nu-2 m^2-5m-2)}{16}\psi_{m+1,\nu}-\cr
\hspace{-1cm}\left.\left.\frac{(\nu+1)(3\nu+2m+4)}{8}\psi_{m+1,\nu+1}+\frac{(\nu+1)(\nu+2)}{16}\psi_{m+1,\nu+2})\right)+{\cal O}\left(\frac{1}{\bar{c}^3}\right)\right],}
\end{eqnarray}
\begin{eqnarray}\label{dpsi_dx-idpsi_dy}
\eqalign{\left(\frac{\partial}{\partial x}-i\frac{\partial}{\partial y}\right)\psi_{m\nu}=\frac{2\sqrt{\bar{c}}}{d}\left[(\nu+m)\psi_{m-1,\nu}+(\nu+1)\psi_{m-1,\nu+1}-{\color{White}\frac{0}{1}}\right.\cr
\frac{1}{\bar{c}}\left(\frac{(\nu+m)(\nu+m-1)}{4}\psi_{m-1,\nu-1}+\frac{(\nu+m)^2}{4}\psi_{m-1,\nu}+\frac{(\nu+1)^2}{4}\psi_{m-1,\nu+1}+\right.\cr
\hspace{-5mm}\left.\frac{(\nu+1)(\nu+2)}{4}\psi_{m-1,\nu+2}\right)+\frac{1}{\bar{c}^2}\left(\frac{(\nu+m)(\nu+m-1)(\nu+m-2)}{16}\psi_{m-1,\nu-2}-\right.\cr
\fl\frac{\nu(\nu+m)(\nu+m-1)}{4}\psi_{m-1,\nu-1}+\frac{(\nu+m)(2\nu^3+2m\nu^2+\nu^2-2m\nu+3\nu-m^2)}{16}\psi_{m-1,\nu}-\hspace{1cm}\cr
\frac{(\nu+1)(2\nu^3+4m\nu^2+5\nu^2+2m^2\nu+4m\nu+7\nu+3m+4)}{16}\psi_{m-1,\nu+1}-\cr
\hspace{-15mm}\left.\left.\frac{(\nu+1)(\nu+2)(\nu+m+1)}{4}\psi_{m-1,\nu+2}+\frac{(\nu+1)(\nu+2)(\nu+3)}{16}\psi_{m-1,\nu+3}\right)+{\cal O}\left(\frac{1}{\bar{c}^3}\right)\right],}
\end{eqnarray}
and
\begin{eqnarray}\label{dpsi_dz}
\frac{\partial}{\partial z}\psi_{m\nu}=\frac{2i}{d}\left[\bar{c}\psi_{m\nu}-\frac{\nu+m}{2}\psi_{m,\nu-1}-\frac{2\nu+m+1}{2}\psi_{m\nu}-\frac{\nu+1}{2}\psi_{m,\nu+1}\right.+\cr
\fl\left.\frac{1}{\bar{c}}\left(\frac{(\nu+m)(\nu+m-1)}{8}\psi_{m,\nu-2}-\frac{\nu(\nu+m)}{4}\psi_{m,\nu-1}-\frac{6\nu^2+6m\nu+6\nu+m^2+3m+2}{8}\psi_{m\nu}-\right.\right.\hspace{8mm}\cr
\left.\left.\frac{(\nu+1)(\nu+m+1)}{4}\psi_{m,\nu+1}+\frac{(\nu+1)(\nu+2)}{8}\psi_{m,\nu+2}\right)+{\cal O}\left(\frac{1}{\bar{c}^2}\right)\right].
\end{eqnarray}
These expressions, Eqs.~(\ref{dpsi_dx+idpsi_dy}-\ref{dpsi_dz}), are extremely powerful. They allow derivatives of the scalar spheroidal wave functions to be evaluated with complete disregard to the actual structure of the $\psi_{m\nu}$. As an example, using Eqs.~(\ref{dpsi_dx+idpsi_dy}-\ref{dpsi_dz}), one can easily check that $\psi_{m\nu}$ indeed satisfies the wave equation.

\section{Vector spheroidal wave functions}\label{section vector wave function}
Defining a set of vector spheroidal wave functions in terms of their Cartesian components is now relatively simple. In analogy to the Gauss-Laguerre solutions to the paraxial approximation being the transverse component of the electric field, we set $E_x$ and $E_y$ as being proportional to a single $\psi_{m\nu}$ and determine $E_z$ using Eqs.~(\ref{dpsi_dx+idpsi_dy}-\ref{dpsi_dz}) such that $\nabla\cdot\mathbf{E}=0$ is satisfied. Due to their $e^{im\phi}$ $\phi$-dependence, the $\psi_{m\nu}$ are eigenfunctions of the angular momentum operator about the $z$-axis. We therefore attempt to define the vector functions so that they are also eigenfunctions of the angular momentum operator about the $z$-axis. This is accomplished by choosing the transverse vector character of the vector wave functions as $\mathbf{\hat{x}}+i\mathbf{\hat{y}}$ and $\mathbf{\hat{x}}-i\mathbf{\hat{y}}$, which corresponds to $\sigma^+$ and $\sigma^-$ light, respectively. We obtain
\begin{eqnarray}\label{E_sigma+}
\eqalign{\mathbf{E}_{J\sigma^+\nu}^+=(\mathbf{\hat{x}}+i\mathbf{\hat{y}})\psi_{J-1,\nu}-\frac{i\mathbf{\hat{z}}}{\sqrt{\bar{c}}}\left[\psi_{J,\nu-1}+\psi_{J\nu}+{\color{White}\frac{0}{1}}\right.\cr
\frac{1}{\bar{c}}\left(\frac{\nu+J-1}{4}\psi_{J,\nu-2}+\frac{7\nu+4J-2}{4}\psi_{J,\nu-1}+\right.\cr
\left.\left.\frac{7\nu+3J+2}{4}\psi_{J\nu}+\frac{\nu+1}{4}\psi_{J,\nu+1}\right)+{\cal O}\left(\frac{1}{\bar{c}^2}\right)\right],}
\end{eqnarray}
\begin{eqnarray}\label{E_sigma-}
\eqalign{\mathbf{E}_{J\sigma^-\nu}^+=(\mathbf{\hat{x}}-i\mathbf{\hat{y}})\psi_{J+1,\nu}+\frac{i\mathbf{\hat{z}}}{\sqrt{\bar{c}}}\left[(\nu+J+1)\psi_{J\nu}+(\nu+1)\psi_{J,\nu+1}+{\color{White}\frac{0}{1}}\right.\cr
\frac{1}{\bar{c}}\left(\frac{(\nu+J)(\nu+J+1)}{4}\psi_{J,\nu-1}+\frac{(\nu+J+1)(5\nu+J+3)}{4}\psi_{J\nu}+\right.\cr
\left.\left.\frac{(\nu+1)(5\nu+4J+7)}{4}\psi_{J,\nu+1}+\frac{(\nu+1)(\nu+2)}{4}\psi_{J,\nu+2}\right)+{\cal O}\left(\frac{1}{\bar{c}^2}\right)\right].}
\end{eqnarray}
We use $J$ as the total angular momentum, equal to the sum of the orbital angular momentum and the spin angular momentum.

So far we have ignored the fact that our solutions to the wave equation are traveling waves which can travel in both the $+\xi$ and $-\xi$ direction. All solutions so far have consistently represented waves traveling in the $+\xi$ direction when an $e^{-i\omega t}$ time dependence is assumed. This is acknowledged in Eqs.~(\ref{E_sigma+}) and (\ref{E_sigma-}) by the $+$ superscript label in $\mathbf{E}_{J\sigma\nu}^+$. We can also define vector spheroidal wave functions $\mathbf{E}_{J\sigma\nu}^-$ traveling in the $-\xi$ direction. These are given by the same expressions, Eqs.~(\ref{E_sigma+}) and (\ref{E_sigma-}), except that $\mathbf{\hat{z}}$ is replaced by $-\mathbf{\hat{z}}$ and the scalar spheroidal wave functions are evaluated at $-\xi$ instead of $\xi$.

For many problems in physics, knowledge of both the electric as well as the magnetic field of an electromagnetic beam is needed~\cite{Jackson98}. Obtaining the magnetic field from the electric field requires taking the curl of the electric field and vice versa. The curl of the vector spheroidal wave functions can be obtained rather elegantly using Eqs.~(\ref{dpsi_dx+idpsi_dy}-\ref{dpsi_dz}) to evaluate the necessary derivatives. We present the result here as an example for the usefulness of these equations.

For an electric field divided into components according to
\begin{equation}
\mathbf{E}=(\mathbf{\hat{x}}+i\mathbf{\hat{y}})E_++(\mathbf{\hat{x}}-i\mathbf{\hat{y}})E_-+\mathbf{\hat{z}}E_z,
\end{equation}
the curl is given by
\begin{eqnarray}
\fl\nabla\times\mathbf{E}=\left[\frac{1}{i}\frac{\partial E_+}{\partial z}+\frac{i}{2}\left(\frac{\partial}{\partial x}-i\frac{\partial}{\partial y}\right)E_z\right](\mathbf{\hat{x}}+i\mathbf{\hat{y}})+\left[i\frac{\partial E_-}{\partial z}+\frac{1}{2i}\left(\frac{\partial}{\partial x}+i\frac{\partial}{\partial y}\right)E_z\right](\mathbf{\hat{x}}-i\mathbf{\hat{y}})+\cr
\left[i\left(\frac{\partial}{\partial x}+i\frac{\partial}{\partial y}\right)E_++\frac{1}{i}\left(\frac{\partial}{\partial x}-i\frac{\partial}{\partial y}\right)E_-\right]\mathbf{\hat{z}}.
\end{eqnarray}
Inserting the necessary expressions from Eqs.~(\ref{dpsi_dx+idpsi_dy}-\ref{E_sigma-}) and writing the result in terms of the $\mathbf{E}_{J\sigma\nu}^+$ we obtain
\begin{eqnarray}\label{rotEsigma+}
\eqalign{\nabla\times\mathbf{E}_{J\sigma^+\nu}^+=\cr
\fl k\left[\mathbf{E}_{J\sigma^+\nu}^++\frac{1}{\bar{c}^2}\left(\frac{(\nu+J-1)(\nu+J-2)}{8}\mathbf{E}_{J\sigma^+,\nu-2}^++\frac{(\nu+J-1)(2\nu+J-1)}{4}\mathbf{E}_{J\sigma^+,\nu-1}^++\right.\right.\cr
\fl\left.\frac{6\nu^2+6J\nu+J^2+J}{8}\mathbf{E}_{J\sigma^+\nu}^++\frac{(\nu+1)(2\nu+J+1)}{4}\mathbf{E}_{J\sigma^+,\nu+1}^++\frac{(\nu+1)(\nu+2)}{8}\mathbf{E}_{J\sigma^+,\nu+2}^+\right)+\hspace{1cm}\cr
\frac{1}{2\bar{c}}(\mathbf{E}_{J\sigma^-,\nu-2}^++2\mathbf{E}_{J\sigma^-,\nu-1}^++\mathbf{E}_{J\sigma^-\nu}^+)+\cr
\hspace{-15mm}\left.\frac{1}{\bar{c}^2}\left(\frac{2\nu+J-1}{2}\mathbf{E}_{J\sigma^-,\nu-2}^++(2\nu+J)\mathbf{E}_{J\sigma^-,\nu-1}^++\frac{2\nu+J+1}{2}\mathbf{E}_{J\sigma^-\nu}^+\right)+{\cal O}\left(\frac{1}{\bar{c}^3}\right)\right]}
\end{eqnarray}
and
\begin{eqnarray}\label{rotEsigma-}
\fl\nabla\times\mathbf{E}_{J\sigma^-\nu}^+=\cr
\fl-k\left[\mathbf{E}_{J\sigma^-\nu}^++\frac{1}{\bar{c}^2}\left(\frac{(\nu+J)(\nu+J+1)}{8}\mathbf{E}_{J\sigma^-,\nu-2}^++\frac{(\nu+J+1)(2\nu+J+1)}{4}\mathbf{E}_{J\sigma^-,\nu-1}^++\right.\right.\cr
\fl\hspace{-5mm}\left.\frac{6\nu^2+6J\nu+12\nu+J^2+5J+6}{8}\mathbf{E}_{J\sigma^-\nu}^++\frac{(\nu+1)(2\nu+J+3)}{4}\mathbf{E}_{J\sigma^-,\nu+1}^++\frac{(\nu+1)(\nu+2)}{8}\mathbf{E}_{J\sigma^-,\nu+2}^+\right)+\cr
\fl\frac{1}{2\bar{c}}[(\nu+J)(\nu+J+1)\mathbf{E}_{J\sigma^+\nu}^++2(\nu+1)(\nu+J+1)\mathbf{E}_{J\sigma^+,\nu+1}^++(\nu+1)(\nu+2)\mathbf{E}_{J\sigma^+,\nu+2}^+]+\cr
\fl\frac{1}{2\bar{c}^2}\left(\frac{(\nu+J)(\nu+J+1)(2\nu+1)}{2}\mathbf{E}_{J\sigma^+\nu}^++(\nu+1)(\nu+J+1)(2\nu+J+2)\mathbf{E}_{J\sigma^+,\nu+1}^++\right.\cr
\fl\left.\left.\frac{(\nu+1)(\nu+2)(2\nu+2J+3)}{2}\mathbf{E}_{J\sigma^+,\nu+2}^+\right)+{\cal O}\left(\frac{1}{\bar{c}^3}\right)\right].
\end{eqnarray}
For $\mathbf{E}_{J\sigma\nu}^-$, the right hand side of Eq.~(\ref{rotEsigma+}) and  Eq.~(\ref{rotEsigma-}) must be multiplied by $-1$ and $\mathbf{E}_{J\sigma\nu}^+$ must be replaced everywhere by $\mathbf{E}_{J\sigma\nu}^-$.

\section{Outlook}
The results obtained here are an excellent starting point for calculating corrections to the paraxial approximation. Due to the equivalence to lowest order in $\frac{1}{\bar{c}}$ of the Laguerre-Gaussian solutions to the paraxial approximation and the spheroidal wave functions considered here, solutions to physical problems based on the paraxial approximation are effectively lowest-order solutions in $\frac{1}{\bar{c}}$ using spheroidal coordinates. Higher-order terms in $\frac{1}{\bar{c}}$ can then be included as perturbations to obtain results to any desired degree of accuracy, provided $\bar{c}$ is large enough that the resulting series converge. This procedure can be used in particular to calculate corrections to the eigenfrequencies and eigenmodes of a Fabry-Perot resonator. The resulting corrections can be observed experimentally in a high-finesse resonator~\cite{Zeppenfeld09}.

A similar procedure as the one used here to write the radial functions in a new form could possibly be applied for solutions to the wave equation in other coordinate systems, in particular for elliptic coordinates. Such solutions should reduce to the Hermite-Gaussian solutions to the paraxial wave equation in the short wavelength limit. Applying such solutions to calculate eigenfrequencies and eigenmodes of two-dimensional resonators would allow comparison with results obtained previously for two-dimensional resonators using other methods~\cite{Lazutkin68,Laabs99,Zomer07}.

\ack
Many thanks to Pepijn W.H. Pinkse for double-checking the calculations. Support from the Deutsche Forschungsgemeinschaft via EUROQUAM (Cavity-Mediated Molecular Cooling) and through the excellence cluster "Munich Centre for Advanced Photonics" is gratefully acknowledged.

\appendix
\section{Recursion relations for Laguerre polynomials~\cite{Abramowitz+Stegun}}\label{appendix recursion}
\begin{equation}
\hspace{-1cm}L_\nu^{(m)}(x)=L_\nu^{(m+1)}(x)-L_{\nu-1}^{(m+1)}(x)
\end{equation}
\begin{equation}
\hspace{-1cm}xL_\nu^{(m)}(x)=-(\nu+m)L_{\nu-1}^{(m)}(x)+(2\nu+m+1)L_\nu^{(m)}(x)-(\nu+1)L_{\nu+1}^{(m)}(x)
\end{equation}
\begin{equation}
\hspace{-1cm}xL_\nu^{(m)}(x)=(\nu+m)L_\nu^{(m-1)}(x)-(\nu+1)L_{\nu+1}^{(m-1)}(x)
\end{equation}
\begin{equation}
\hspace{-1cm}x\frac{d}{dx}L_\nu^{(m)}(x)=\nu L_\nu^{(m)}(x)-(\nu+m)L_{\nu-1}^{(m)}(x)
\end{equation}

\footnotesize
\section{$r_{m\nu}^{(p)}$ for $p=3$ and $p=4$}\label{appendix r_mnu^p}
\begin{eqnarray}
\eqalign{r_{m\nu}^{(3)}(\xi)=\frac{1}{192}\left(\frac{4(\nu+1)(\nu+2)(\nu+3)(\nu+m+1)(\nu+m+2)(\nu+m+3)}{(1+i\xi)^3}-\right.\cr
\frac{4\nu(\nu-1)(\nu-2)(\nu+m)(\nu+m-1)(\nu+m-2)}{(1-i\xi)^3}-\cr
\frac{6(\nu+1)(\nu+2)(\nu+m+1)(\nu+m+2)(\nu^2+m\nu-4\nu-2m-2)}{(1+i\xi)^2}+\cr
\frac{6\nu(\nu-1)(\nu+m)(\nu+m-1)(\nu^2+m\nu+6\nu+3m+3)}{(1-i\xi)^2}-\cr
\frac{3(\nu+1)(\nu+m+1)(14\nu^3+21m\nu^2-13\nu^2+7m^2\nu-13m\nu-24\nu-4m^2-12m-8)}{1+i\xi}-\cr
\left.\frac{3\nu(\nu+m)(14\nu^3+21m\nu^2+55\nu^2+7m^2\nu+55m\nu+44\nu+11m^2+22m+11)}{1-i\xi}\right)}
\end{eqnarray}
\\\\
\begin{eqnarray}
\eqalign{\fl r_{m\nu}^{(4)}(\xi)=\frac{1}{768}\left(\frac{2(\nu+1)(\nu+2)(\nu+3)(\nu+4)(\nu+m+1)(\nu+m+2)(\nu+m+3)(\nu+m+4)}{(1+i\xi)^4}+\right.\\
\frac{2\nu(\nu-1)(\nu-2)(\nu-3)(\nu+m)(\nu+m-1)(\nu+m-2)(\nu+m-3)}{(1-i\xi)^4}-\\
\frac{4(\nu+1)(\nu+2)(\nu+3)(\nu+m+1)(\nu+m+2)(\nu+m+3)(\nu^2+m\nu-6\nu-3m-3)}{(1+i\xi)^3}-\\
\frac{4\nu(\nu-1)(\nu-2)(\nu+m)(\nu+m-1)(\nu+m-2)(\nu^2+m\nu+8\nu+4m+4)}{(1-i\xi)^3}+\\
\frac{(\nu+1)(\nu+2)(\nu+m+1)(\nu+m+2)}{(1+i\xi)^2}\times\\
\fl(\nu^4+2m\nu^3-54\nu^3+m^2\nu^2-81m\nu^2+111\nu^2-27m^2\nu+111m\nu+180\nu+30m^2+90m+60)+\hspace{2cm}\\
\frac{\nu(\nu-1)(\nu+m)(\nu+m-1)}{(1-i\xi)^2}\times\\
\fl(\nu^4+2m\nu^3+58\nu^3+m^2\nu^2+87m\nu^2+279\nu^2+29m^2\nu+279m\nu+208\nu+58m^2+104m+46)+\\
\fl\frac{(\nu+1)(\nu+m+1)}{1+i\xi}\left(\nu^6+(3m+2)\nu^5+(3m^2+5m-311)\nu^4+(m^3+4m^2-622m+84)\nu^3+\right.\\
\fl\left.(m^3-382m^2+126m+653)\nu^2+(-71m^3+102m^2+653m+540)\nu+30m^3+162m^2+270m+138\right)+\\
\fl\frac{\nu(\nu+m)}{1-i\xi}\left(\nu^6+(3m+4)\nu^5+(3m^2+10m-306)\nu^4+(m^3+8m^2-612m-1328)\nu^3+2m^3\nu^2-\right.\\
\fl\hspace{-5mm}\left.{\color{White}\frac{0}{1}}\left.(376m^2+1992m+1470)\nu^2-(70m^3+866m^2+1470m+734)\nu-(101m^3+323m^2+367m+145)\right)\right)}
\end{eqnarray}

\bibliographystyle{unsrt}

\end{document}